\documentclass[12pt]{article}
\usepackage{graphicx}
\usepackage{epsfig}
\usepackage{rotating} 
\makeindex

\usepackage{relsize}
\def\babar{\mbox{\slshape B\kern-0.1em{\smaller A}\kern-0.1em
B\kern-0.1em{\smaller A\kern-0.2em R}}}
\begin{document}
\pagestyle{plain}

\def\cerenkov{$\check{\rm C}{\rm erenkov}$ }

\begin{center}
\Large {\bf ISR physics at \babar\ }
\end{center}
\smallskip
\renewcommand{\thefootnote}{\fnsymbol{footnote}}
\begin{center}
{\large
S.Serednyakov\footnote[1]{e-mail: seredn@inp.nsk.su}, 
}\\

\renewcommand{\thefootnote}{\arabic{footnote}}
\vskip 0.5cm
{\it Budker Institute of Nuclear Physics, Novosibirsk, Russia} \\
\end{center}

\smallskip

\begin{center}
\begin{abstract}
 A method of measuring  e+e- annihilation cross sections at low           
energy $\sqrt{s}<$ 5 GeV, using initial-state radiation, is described.     
Experimental data from the PEP-II B-factory at 10.6 GeV center-of-mass 
energy, obtained    
via ISR, are presented. The cross sections are measured for many           
processes   $e^+e^-\to 3\pi$, $4\pi$, $2K\pi$, $2K2\pi$,  $4K $, $p\bar{p}$, 
$\Lambda\bar{\Lambda}$,  $D\bar{D}$, $\ldots$ . 
From the measured cross sections the parameters of known resonances           
are improved, the baryons form factors are derived and compared with
theory predictions.          
New states, e.g;  Y(4260) and  Y(2175),  for which  the internal structure is not yet
established, are observed. 
 \end{abstract}
\end{center}

\

{\underline{\bf  Physics of the ISR method.}}
Processes of $e^+e^-$ annihilation
  \begin{equation}   
     e^+e^-\to  hadrons,
     \label{eq1}
       \end{equation}
can be studied in a wide  center-of-mass (c.m.) 
energy range  at a high luminosity  $e^+e^-$
machine  using 
initial-state radiation (ISR) \cite{ResNumb} in the reaction

\begin{equation}
     e^+e^-\to hadrons +\gamma,
         \label{eq11}
           \end{equation}
with an energetic recoil photon (ISR photon).  The following equation
\cite{Wxs} relates  cross sections of the processes (\ref{eq1}) and 
(\ref{eq11})

\begin{equation}
  \frac{d\sigma_{e^+e^-\to hadrons + \gamma}(m)}{dm} = 
    \frac{2m}{s}\cdot W(s,x,\theta^*_0)\cdot\sigma (m),
      \label{eq4}
       \end{equation}
where $s$ is the squared c.m. total energy  
and $ E_0={\sqrt s}/2$ is the c.m. beam energy,
$m$ is the invariant mass of the  final state  hadrons,
$x=\frac{E^*_{\gamma}}{E_0}=1-\frac{m^2}{s}$, $E^*_{\gamma}$ and $\theta^*_\gamma$
are the ISR photon c.m. energy and minimal c.m. polar angle 
and $\sigma (m)$ is the $e^+e^-\to  hadrons$ cross section.
The radiation function $W(s,x,\theta^*_0)$ ~\cite{Wxs}

\begin{equation}
W(s,x,\theta^*_0) = \frac{\alpha}{\pi x}\cdot( (2-2x+x^2)\cdot
\ln\frac{1+cos\theta^*_0}{1-cos\theta^*_0} -x^2cos\theta^*_0)
\label{eq5}
\end{equation}
describes the probability of the ISR photon emission, integrated
over the ISR photon polar angle
$\theta^*_\gamma$ in limits $\theta^*_\gamma>\theta^*_0$,
$\alpha$ is  the fine structure constant. In case of 
$\theta^*_0\rightarrow 0$ the function $W(s,x)$ is modified as follows
$W(s,x) = \frac{\alpha}{\pi x}\cdot(L-1) (2-2x+x^2)$ with
$L=2\ln({\sqrt s}/m_e)$, where $m_e$ is an electron mass.

  In recent years many cross section measurements were performed
with the  \babar\ detector \cite{BBNim} at the PEP-II
asymmetric-energy storage ring using the ISR approach.  One can recall, that 
at PEP-II  the 9-GeV electrons collide with
the 3.1-GeV positrons at a c.m.  energy of 10.6~GeV
(the $\Upsilon$(4S) resonance). The measured ISR  hadronic mass 
varied in wide limits from $\sim$ 1 GeV/c$^2$ up to $\sim$ 5 GeV/c$^2$.

    Since  the polar angle distribution of the ISR photon is peaked near
$\theta^*_{\gamma}=0,180^o$, only a small part  $\sim15$\% of ISR photons is detected in
the \babar\ 
calorimeter.  Two modes of ISR measurements are used in \babar\ :
1 - ISR photon is detected in an event and used as the  tag,
2 - ISR photon is emitted close to the beam axis  and its
detection is not required.  In the latter  case the  
recoil mass against  the hadronic system  is expected to be zero. 
Most of  the  ISR measurements 
in \babar\ are performed in the  mode with detection of the ISR photon.
Only a few measurements with  hadronic mass $\simeq 4 GeV/c^2$ used
the mode without detection of the ISR photon.

\begin{figure} 
\begin{minipage}[t]{0.4\textwidth} 
\includegraphics[width=.98\textwidth]{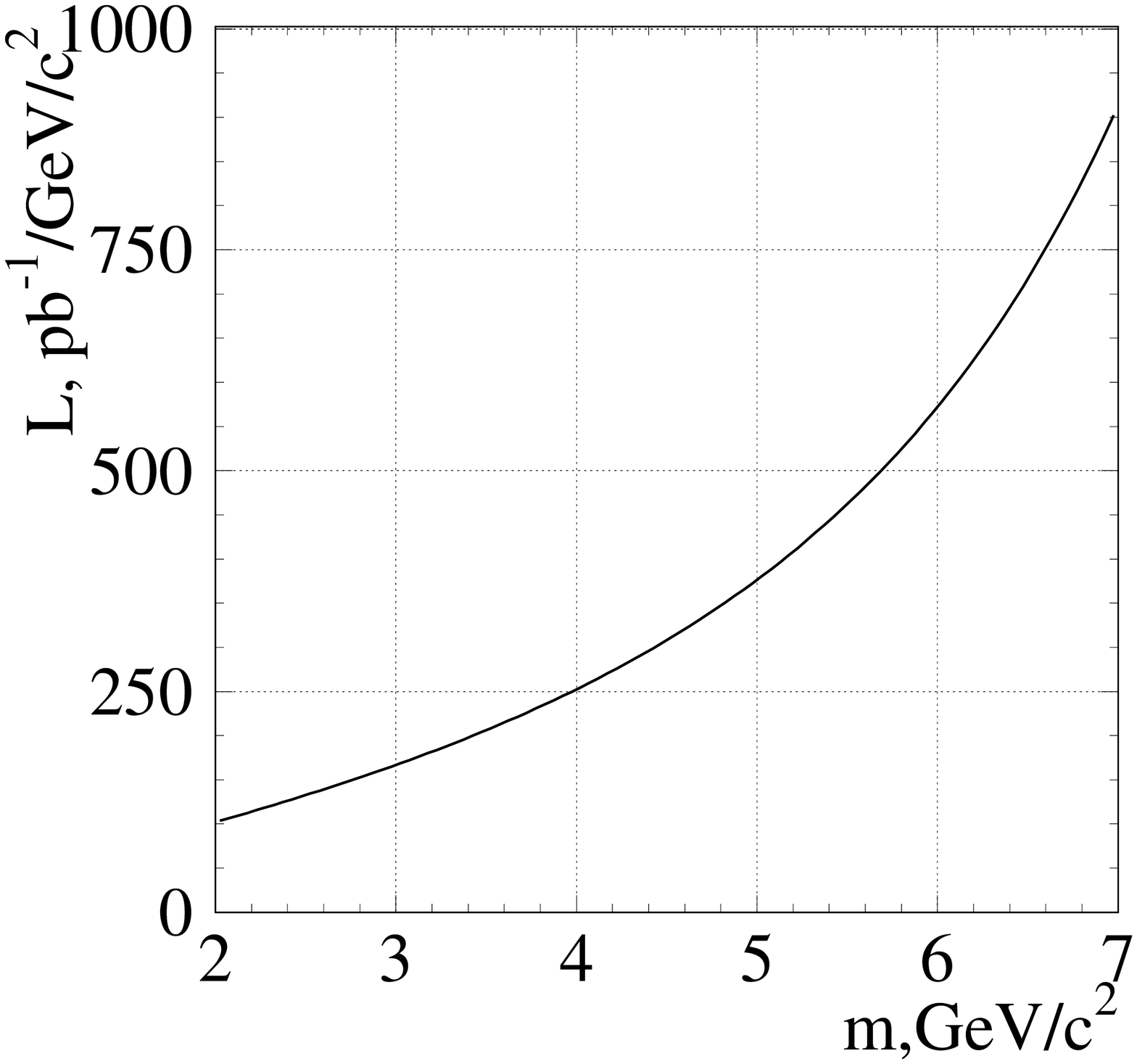} 
\caption{The mass dependence of ISR luminosity, 
calculated for the  \babar\ luminosity  of  454 1/fb  
and the  ISR photon angle range 
 $20^o<\theta^*_{\gamma}<160^0$.}
\label{ISRLum} 
\end{minipage} 
\hfill
\begin{minipage}[t]{0.55\textwidth}
\includegraphics[width=.98\textwidth]{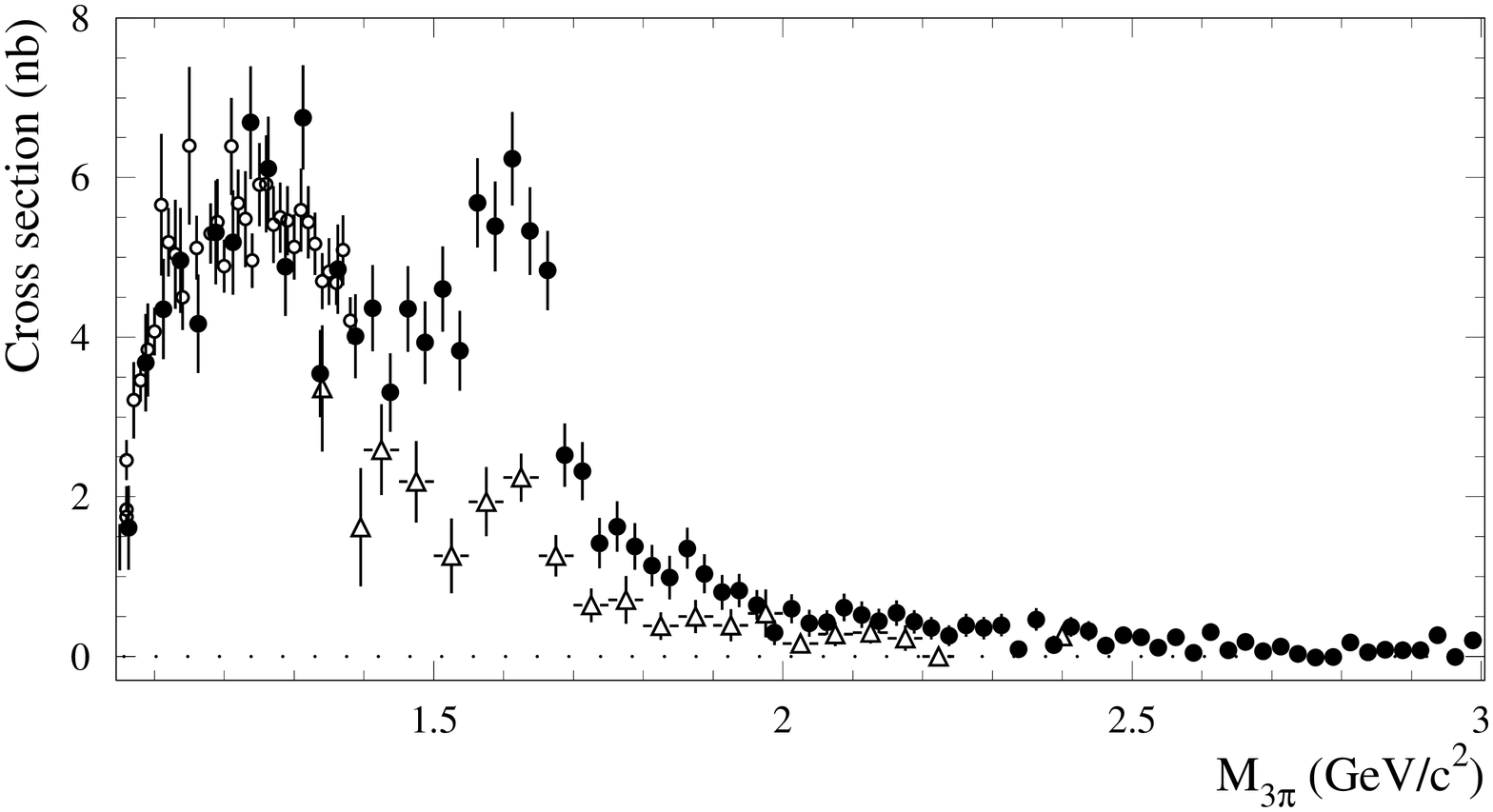}
\caption{The $e^+e^-\to \pi^+\pi^-\pi^0$ cross section, 
measured by SND (open circles) , DM2 (open triangles), 
and \babar\ (filled circles).}
\label{3pi}
\end{minipage}
\begin{minipage}[t]{0.46\textwidth}
\includegraphics[width=.98\textwidth]{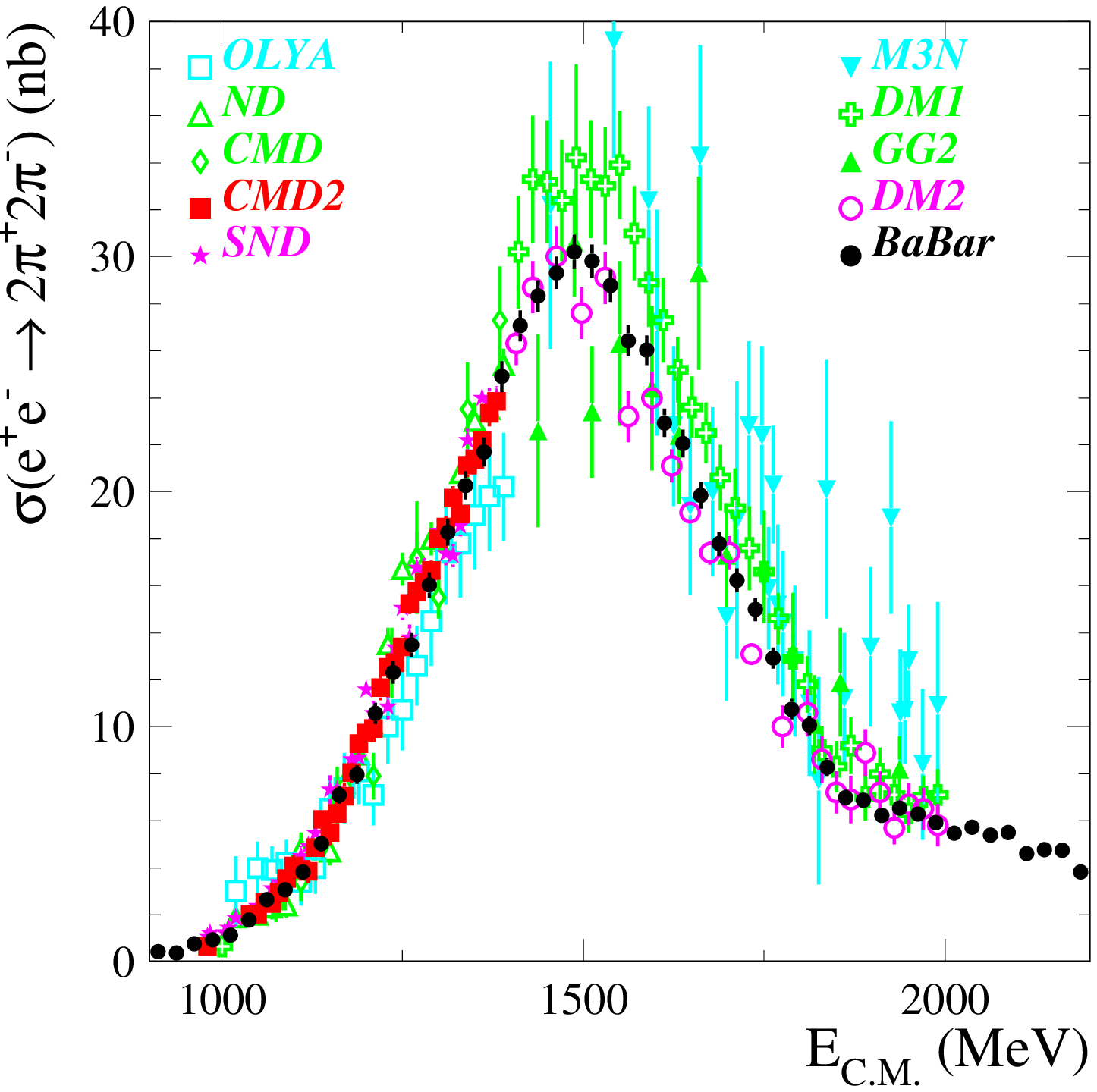}
\caption{The comparison of the 
$e^+e^-\to \pi^+\pi^-\pi^+\pi^-$ cross section,
measured in  \babar\ experiment with the results 
of previous measurements. } 
\label{4pi1}
\end{minipage}
\hfill
\begin{minipage}[t]{0.5\textwidth}
\includegraphics[width=.98\textwidth]{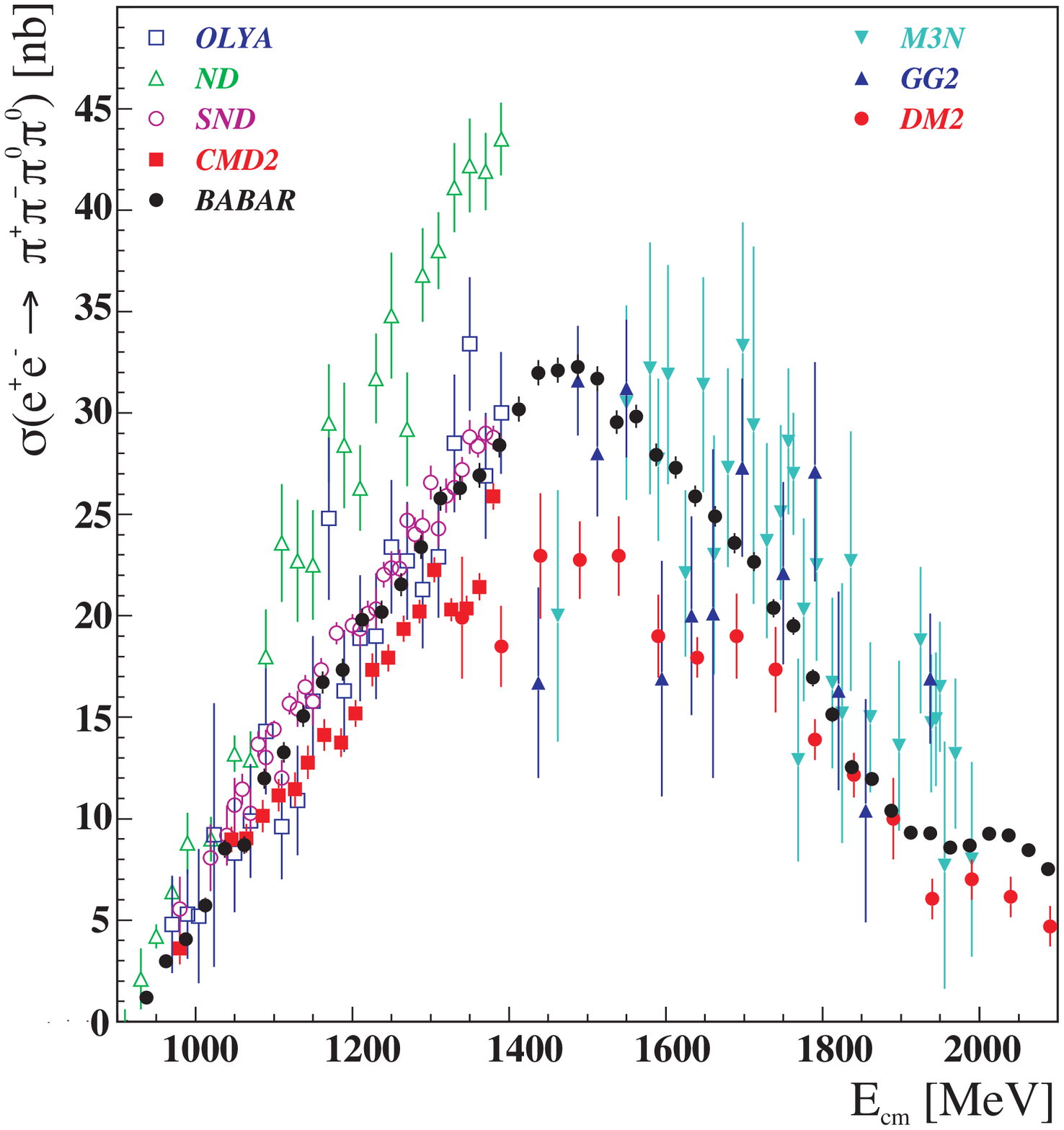}
\caption{The comparison of the 
$e^+e^-\to \pi^+\pi^-\pi^0\pi^0$ cross section,
measured in  \babar\ experiment (preliminary data)  
with the results  of previous measurements.
An  enhancement near 2.1 GeV is seen. }
\label{4pi2}
\end{minipage}
\end{figure}

The cross-section   of particular $e^+e^-$ annihilation 
process is calculated from
the  measured hadronic  mass spectrum using the expression
\begin{equation}
\frac{dN}{dm}~=~ \varepsilon(m)~R(m)~\sigma (m)~\frac{dL}{dm},
\label{crss}
\end{equation}
where $dL/dm$ is the so called ISR differential luminosity,
$\varepsilon$ is the detection efficiency, and  R 
is the radiative correction factor. The ISR luminosity is calculated
using the total integrated luminosity $L_0$ and the probability
of the  ISR photon emission (Eq.~\ref{eq5}):
\begin{equation}
\frac{{\rm d}L}{{\rm d}m}=
W(s,x,\theta^*_0)\frac{2m}{s}\,L_0.
\label{ISRlum}
\end{equation}
In the mode with the detection of the ISR photon in the \babar\
calorimeter we have 
$30^\circ~<~\theta^\star_\gamma~<~150^\circ$
and $\theta^\star_0=30^\circ$. The plot of the ISR luminosity in  \babar\ versus
hadronic invariant mass with L$_0$=454 fb$^{-1} $
is shown in Fig.\ref{ISRLum}.  The ISR  luminosity, integrated over mass
range from 2 to 5 GeV/c$^2$, is rather high -  0.63 fb$^{-1}$.

  The ISR technique allows to utilize 
the high luminosity of the   $e^+e^-$ factories (PEP-II, KEKB, KLOE),
operating at the fixed energy. 
It is worth to compare the ISR with the  direct 
$e^+e^-$ experiments. First of all, the ISR technique allows to study the 
energy range from 
2m$_\pi$c$^2$ up to $\sqrt{s}$ in a single experiment.
 The second ISR advantage is that the detection
efficiency in the mode with detected ISR photon   
is finite at the hadron polar angle $\theta\simeq  0$
(in the hadron system frame). In the direct $e^+e^-$ experiment
the hadrons produced at the very small angles  are not seen at all.
   The next point  is that in ISR the detection efficiency is
finite at the very threshold of the particular reaction, while 
in the conventional   $e^+e^-$ experiment the detection efficiency vanishes
at the threshold because of the low momenta  of produced particles.

     But one should mention that the energy resolution 
and  the energy scale calibration accuracy in ISR are worser than in
the direct $e^+e^-$ set up. 
Also, in  the direct $e^+e^-$ study of 
narrow resonances one could concentrate 
the integrated luminosity in the resonance peak and thus  
have much   higher number of events than in the ISR study. 

 New ISR measurements of $e^+e^-\to hadrons$ cross sections are important
because they  are  used in the calculation of the Standard model parameters
such as the 
muon anomaly $a_{\mu}=\frac{g-2}{2}$ and fine structure constant
at Z-mass $\alpha_{em}(s=M_Z^2)$.
At present, there is the considerable disagreement $\sim 3\sigma$ between
the $ a_{\mu}$ measurement  and   calculation; so new  
accurate cross section data are needed.

In this review the  main  \babar\ ISR results are  presented.
The results are divided into three  parts: production of mesons, 
production of baryons,  and resonance physics.    

{\underline{\bf Production of mesons}}
     The processes of the $e^+e^-$ annihilation into light mesons ($\pi$, K)
give the highest contribution to the total hadronic cross section up to 3 GeV.
Many reactions have been  studied at  \babar\ using ISR. Figure \ref{3pi} shows the 
experimental data on the 
$e^+e^-\to \pi^+\pi^-\pi^0$ reaction \cite{BB3pi}. 
One can see that the 
 \babar\ data well agree with SND measurement 
\cite{SND3pi} below 1.4 GeV and strongly
contradict DM2 results \cite{DM23pi} at higher energy.  
The  \babar\ cross section clearly shows 
two states $\omega(1420)$ and $\omega(1650)$, which are
considered as  the radial and orbital excitations of $\omega (783)$.
    
      The channels $e^+e^-\to \pi^+\pi^-\pi^+\pi^-,\pi^+\pi^-\pi^0\pi^0$ 
give  the  largest contribution to the hadronic vacuum 
polarization term of $a_{\mu}$ above 1 GeV.  
 \babar\ data in Figs.\ref{4pi1},\ref{4pi2} show the 
considerable improvement of the accuracy above 1.4 GeV \cite{BB4pipm,BB4pi00}.   
In these reactions the
$\omega\pi^0$, $\pi a_1$, $\rho^+\rho^-$, $\rho^0 f^0$ 
intermediate states dominate.
In the channel $e^+e^-\to\pi^+\pi^-\pi^0\pi^0$  a peak
above 2 GeV (Fig.\ref{4pi2}) is seen,  which can be manifestation of the 
hypothetical $\rho (2150)$   or $\rho^{\prime\prime\prime}$ state.

\begin{figure}
\begin{minipage}[t]{0.46\textwidth}
\includegraphics[width=.98\textwidth]{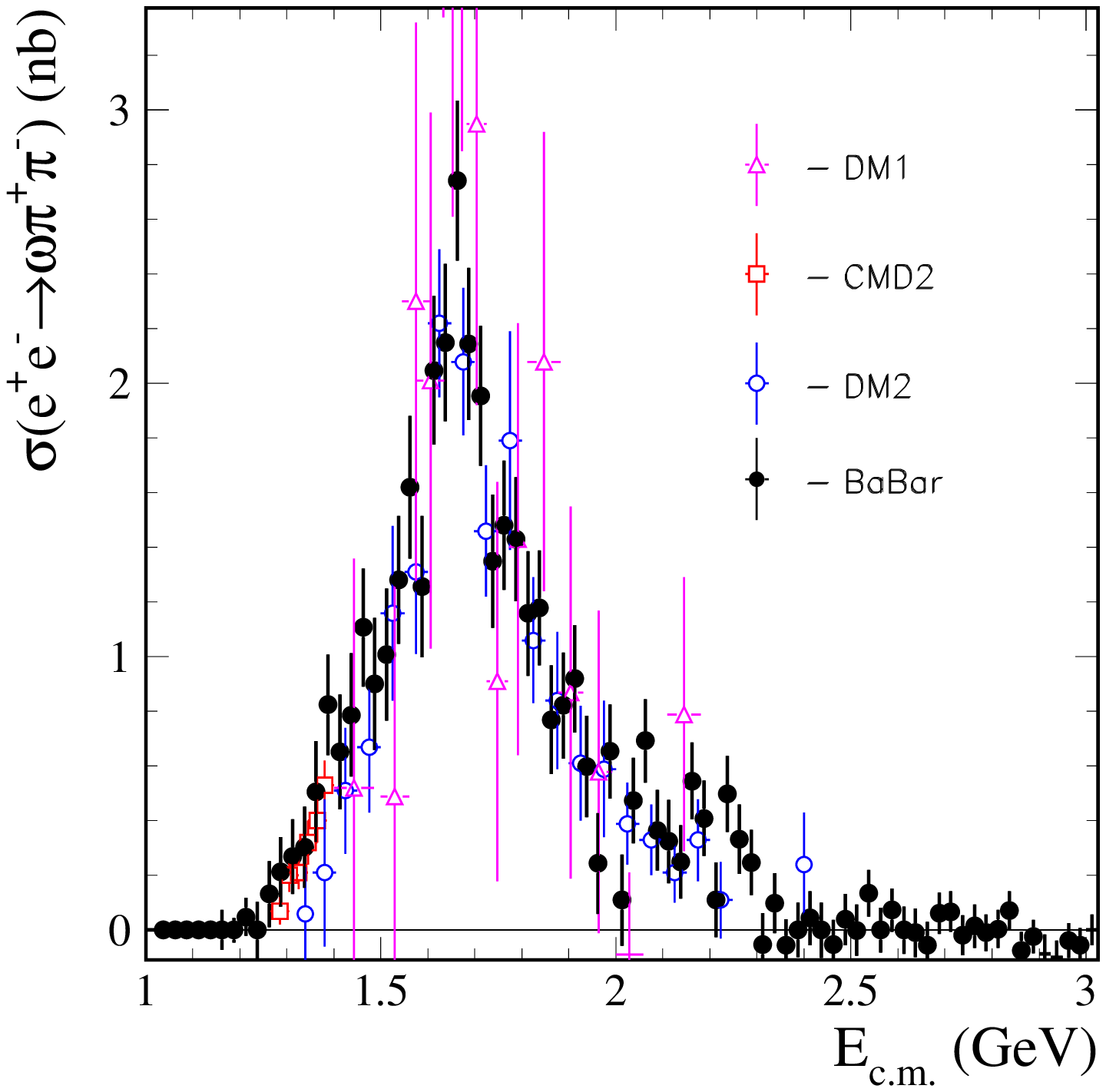}
\caption{The comparison of the 
$e^+e^-\to \omega\pi^+\pi^-$ cross section,
measured in  \babar\ experiment with the results 
of previous measurements. }
\label{om2pi}
\end{minipage}
\hfill
\begin{minipage}[t]{0.46\textwidth}
\includegraphics[width=.98\textwidth]{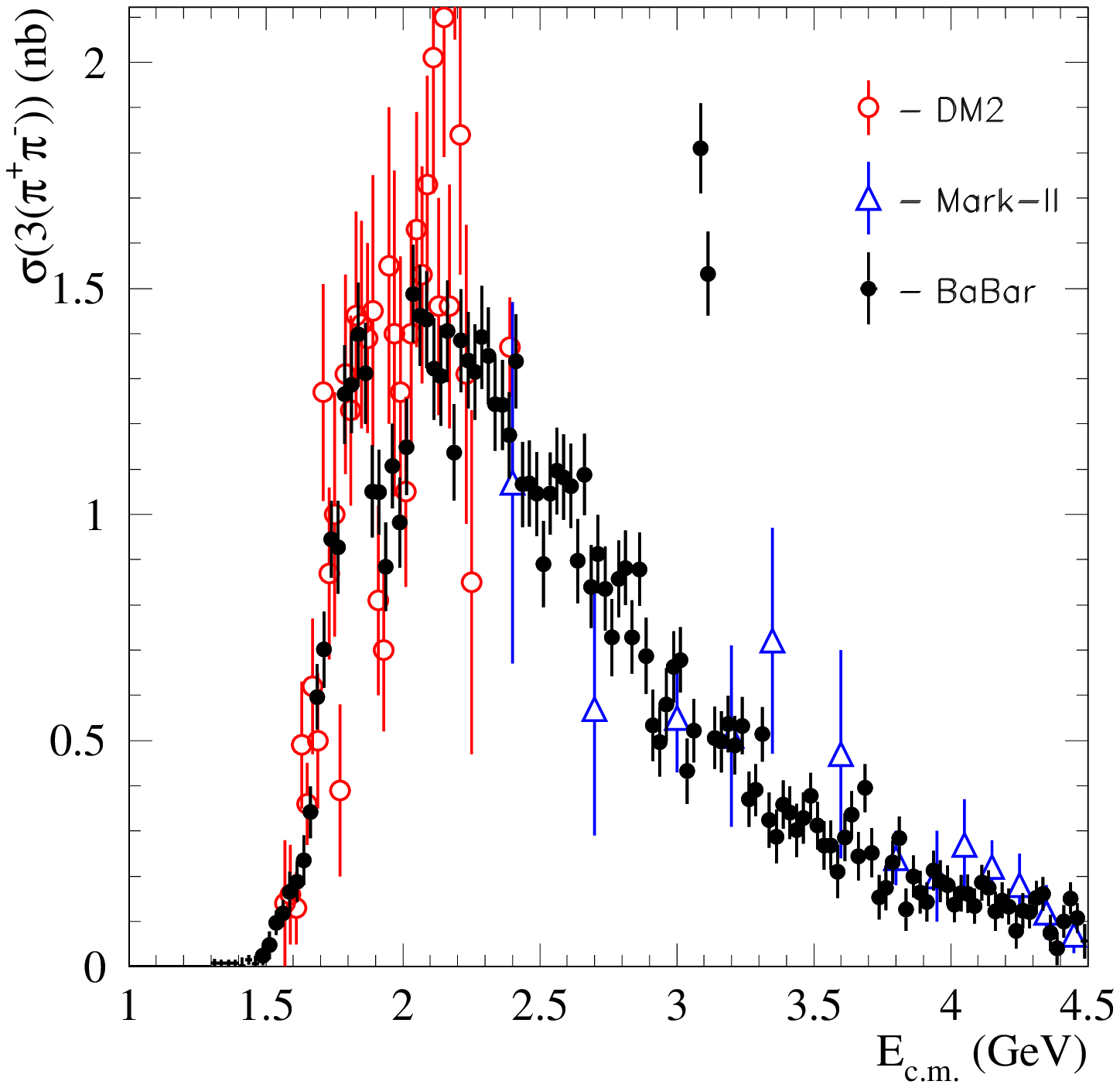}
\caption{ The comparison of the 
$e^+e^-\to 3(\pi^+\pi^-)$ cross section,
measured in  \babar\ experiment with the results 
of previous measurements. A dip near 1.9 GeV
is seen. }
\label{6pi}
\end{minipage}
\begin{minipage}[t]{0.46\textwidth}
\includegraphics[width=.98\textwidth]{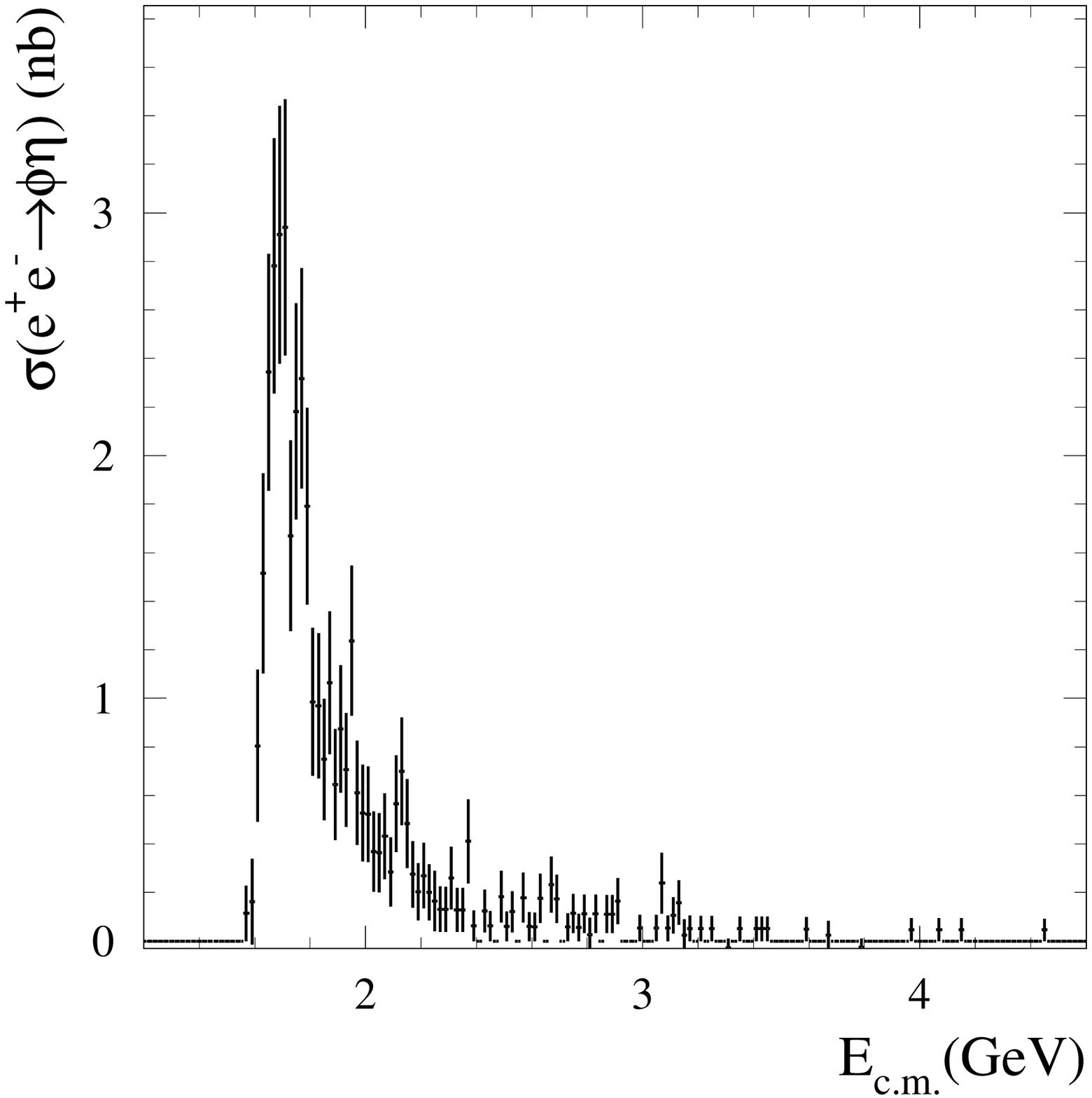}
\caption{The $e^+e^-\to \phi\eta$ cross
section measured in  \babar\ experiment. An enhancement at 2.15
GeV is seen. }
\label{fieta}
\end{minipage}
\hfill
\begin{minipage}[t]{0.46\textwidth}
\includegraphics[width=.98\textwidth]{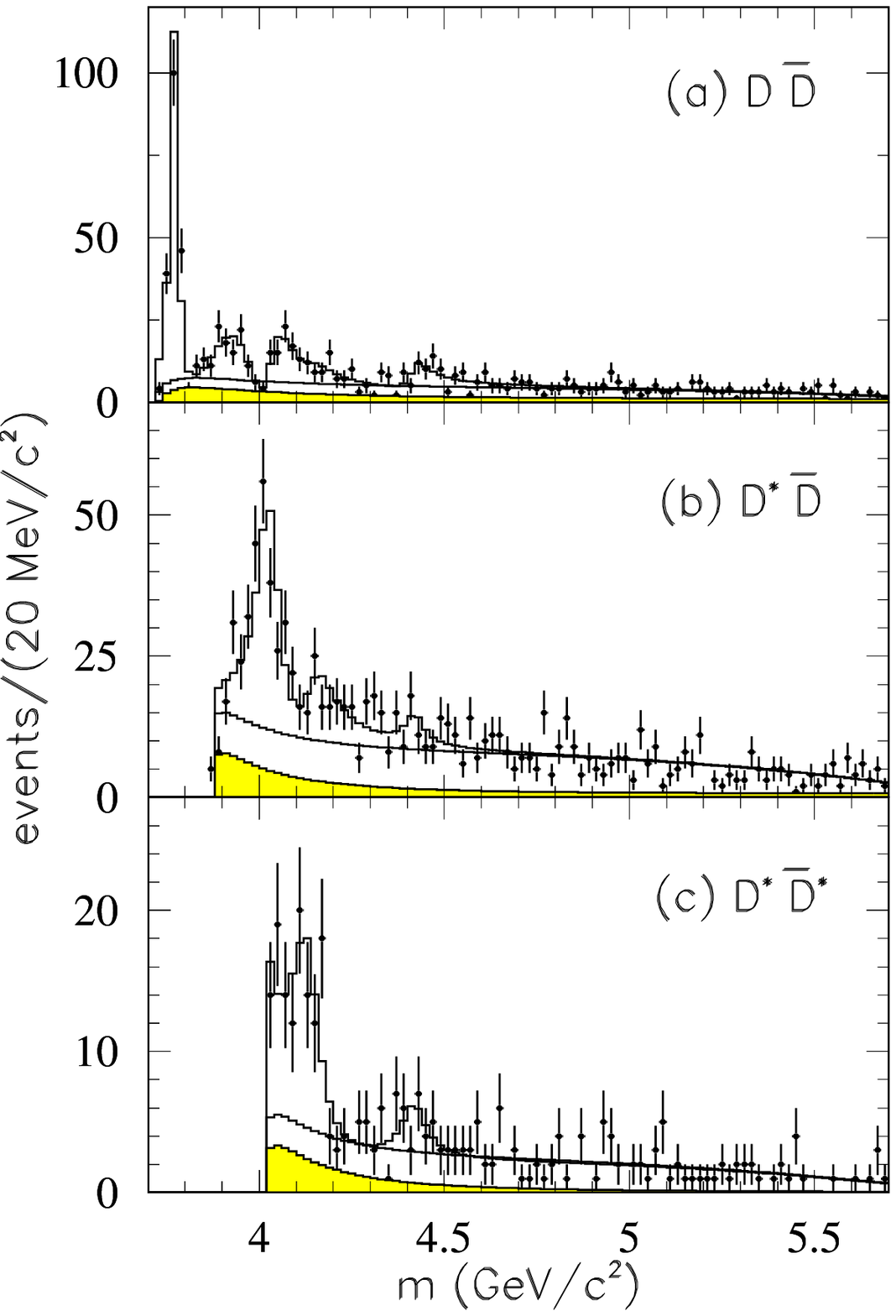}
\caption{The  \babar\  ISR measurements of  the 
$e^+e^-\to D\bar{D}$, $D^*\bar{D}$, 
and $D^*\bar{D^*}$ cross sections. }
\label{DD}
\end{minipage}
\end{figure}

   The cross sections of  $e^+e^-\to 5\pi , 6\pi$ processes \cite{BB5pi,BB6pi} 
have been  
measured by  \babar\ from the threshold up to 4.5 GeV in the
final states $2(\pi^+\pi^-)\pi^0, ~2(\pi^+\pi^-)2\pi^0, ~
3(\pi^+\pi^-)$. The
final states have very complex structure. As an example, the 
$e^+e^-\to \omega\pi^+\pi^-$    cross section 
with the domination of the $\omega (1650)$ state is shown in
Fig.\ref{om2pi}. 
In the $e^+e^-\to 6\pi$ cross section the structure below 2 GeV 
is seen (Fig.\ref{6pi}), which is assumed to  correspond to the state 
with the parameters 
M=1.88$\pm 0.03$ GeV/c$^2$ and $\Gamma$=0.13$\pm$0.03 GeV/c$^2$ 
\cite{BB6pi}.

   The channels with kaons are not less interesting. In the $e^+e^-\to K^+K^-\pi^0$ 
and $e^+e^-\to K^{\pm}K_S\pi^{\mp}$ reactions \cite{BB2Kpi} 
the $\phi (1650)$ isoscalar
resonance dominates. The cross section for the   $\phi\eta$ and $\phi\pi^0$
final states have been  measured for the first time. 
In the   $\phi\eta$ final state
the candidate for the Y(2175) or $\phi^{\prime\prime}$  state is possibly seen
(Fig.\ref{fieta}).  The processes $e^+e^-\to K^+K^-\pi^+\pi^-,
K^+K^-\pi^0\pi^0$ \cite{BB2K2pi}
show domination of the $\phi(1650)$ in the total cross
section and $K^{\star}(890),~K_1(1270),~\eta(1500)$ 
in   intermediate states.
The cross section of rare processes with kaons
$e^+e^-\to K^+K^-\pi^+\pi^-\pi^0, K^+K^-\pi^+\pi^-\eta$ \cite{BB5pi}, 
$e^+e^-\to K^+K^-\pi^+\pi^-\pi^+\pi^-$ \cite{BB6pi}  and  
$e^+e^-\to K^+K^-K^+K^-$ \cite{BB2K2pi} 
have been  also measured by  \babar\ for the first time.
 The maximum cross section value for these processes is 
not higher than 1 nb.

   The $e^+e^-$ annihilation into  D-mesons was studied by  \babar\ 
\cite {BBDD}   in the mode with undetected ISR photon.
The D-mesons ($D^{\pm},D^0,\bar{D^0}$, $D^{*\pm},D^{0*},\bar{D^{0*}}$) 
are selected via  their main decay modes ($K\pi, K2\pi, K3\pi$). 
The main goals of this study are  to confirm the high mass charmonium states
$\psi (4040),\psi (4160),\psi (4400)$ and look for the recently observed
$Y(4260)$ meson in the $D\bar{D}$ decay mode. The measured cross sections are
presented in Figs.\ref{DD}a,b,c  for $D\bar{D}$, $D^*\bar{D}$, and $D^*\bar{D^*}$
modes respectively. While known  charmonium states are seen well, no sign
of $Y(4260)$ is observed; so only upper limits on $Y(4260)$ branching
fractions into  $D\bar{D}$ are set. It has been proved that 
$BF(Y(4260)\to D\bar{D})<BF(Y(4260)\to J/\psi\pi^+\pi^-)$.

\begin{figure}
\begin{minipage}[t]{0.55\textwidth}
\includegraphics[width=.98\textwidth]{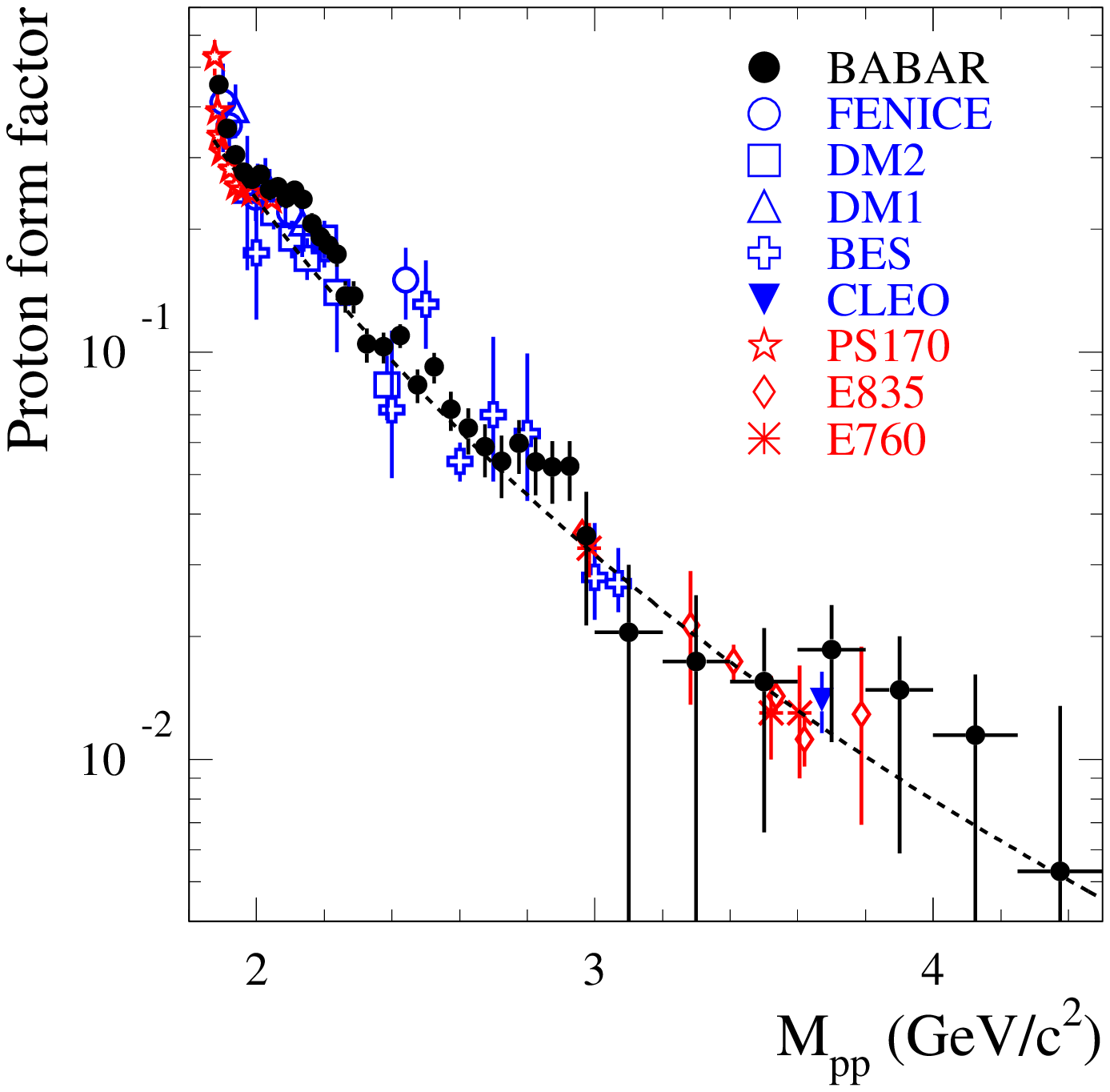}
\caption{The proton form factor.
The solid line corresponds to the  QCD fit \cite{Chern2}.}
\label{prot1}
\end{minipage}
\hfill
\begin{minipage}[t]{0.40\textwidth}
\includegraphics[width=.98\textwidth]{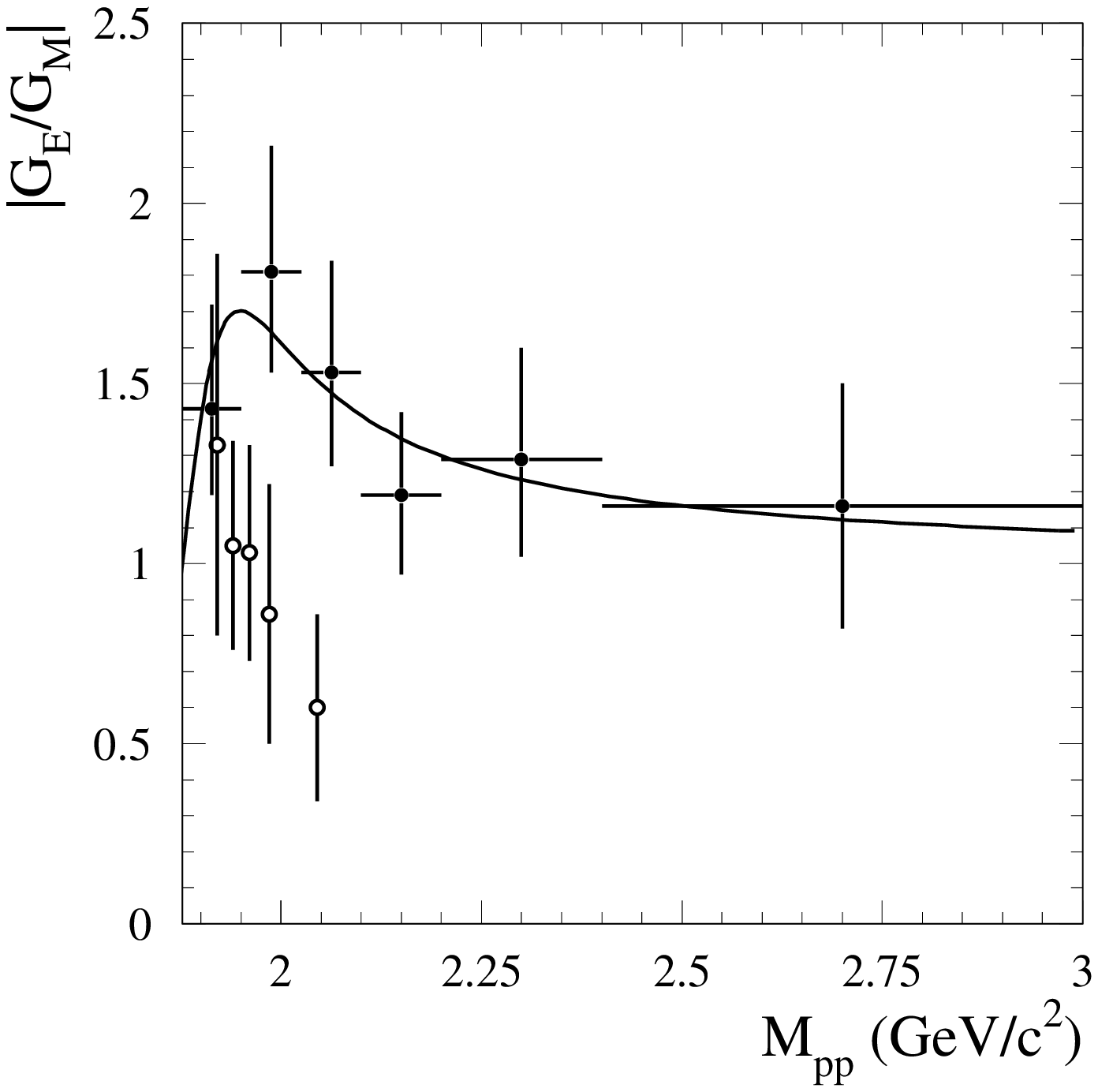}
\caption{The  measured by \babar\  $G_E/G_M$ ratio
(black points),  compared with LEAR data (open circles). }
\label{prot2}
\end{minipage}
\begin{minipage}[t]{0.30\textwidth}
\includegraphics[width=.98\textwidth]{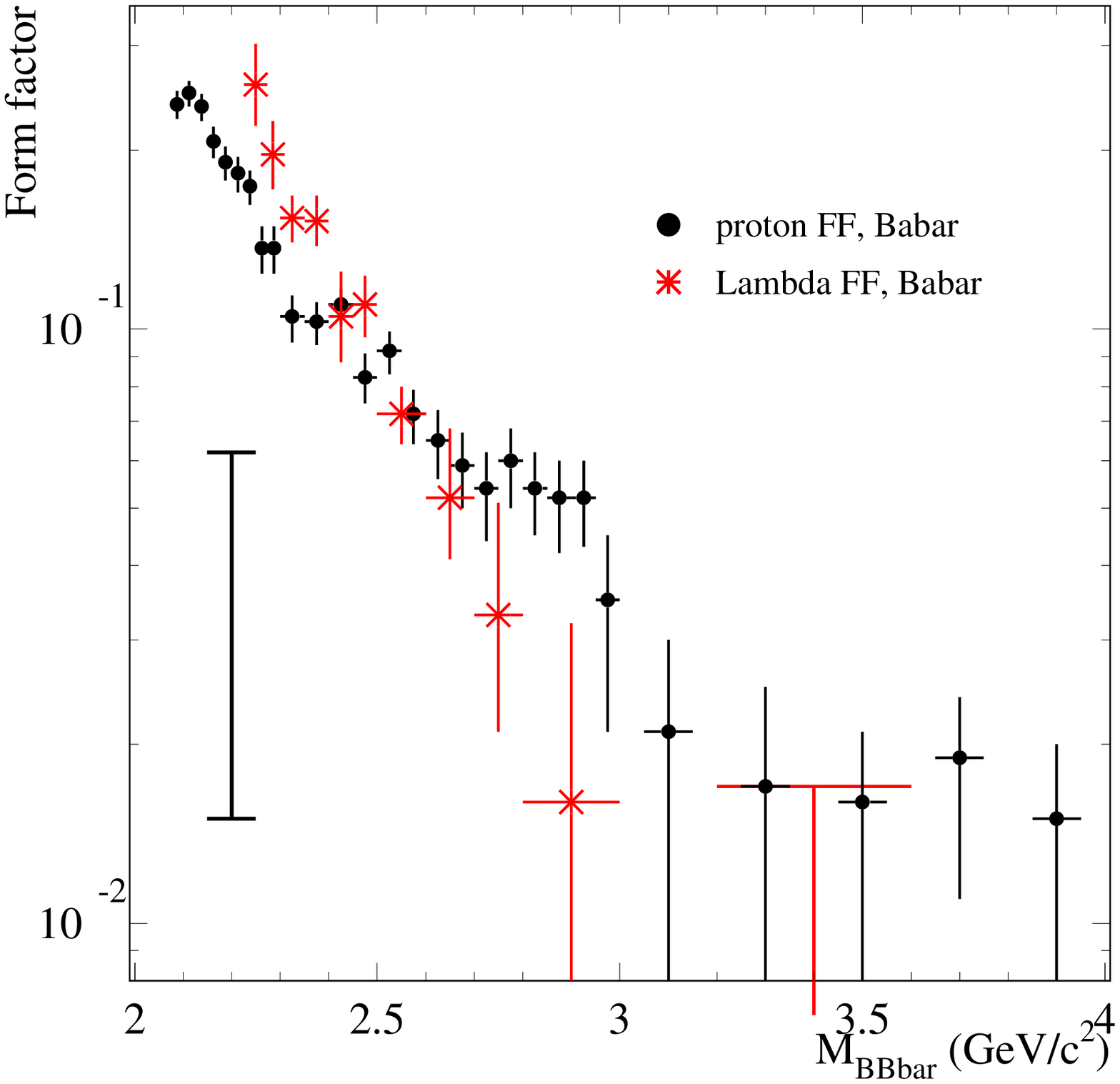}
\caption{The comparison of proton and $\Lambda$ 
form factors. Logarithmic scale. The vertical line corresponds 
to the asymptotic ratio between proton and  $\Lambda$ 
form factors. }
\label{prot3}
\end{minipage}
\hfill
\begin{minipage}[t]{0.35\textwidth}
\includegraphics[width=.98\textwidth]{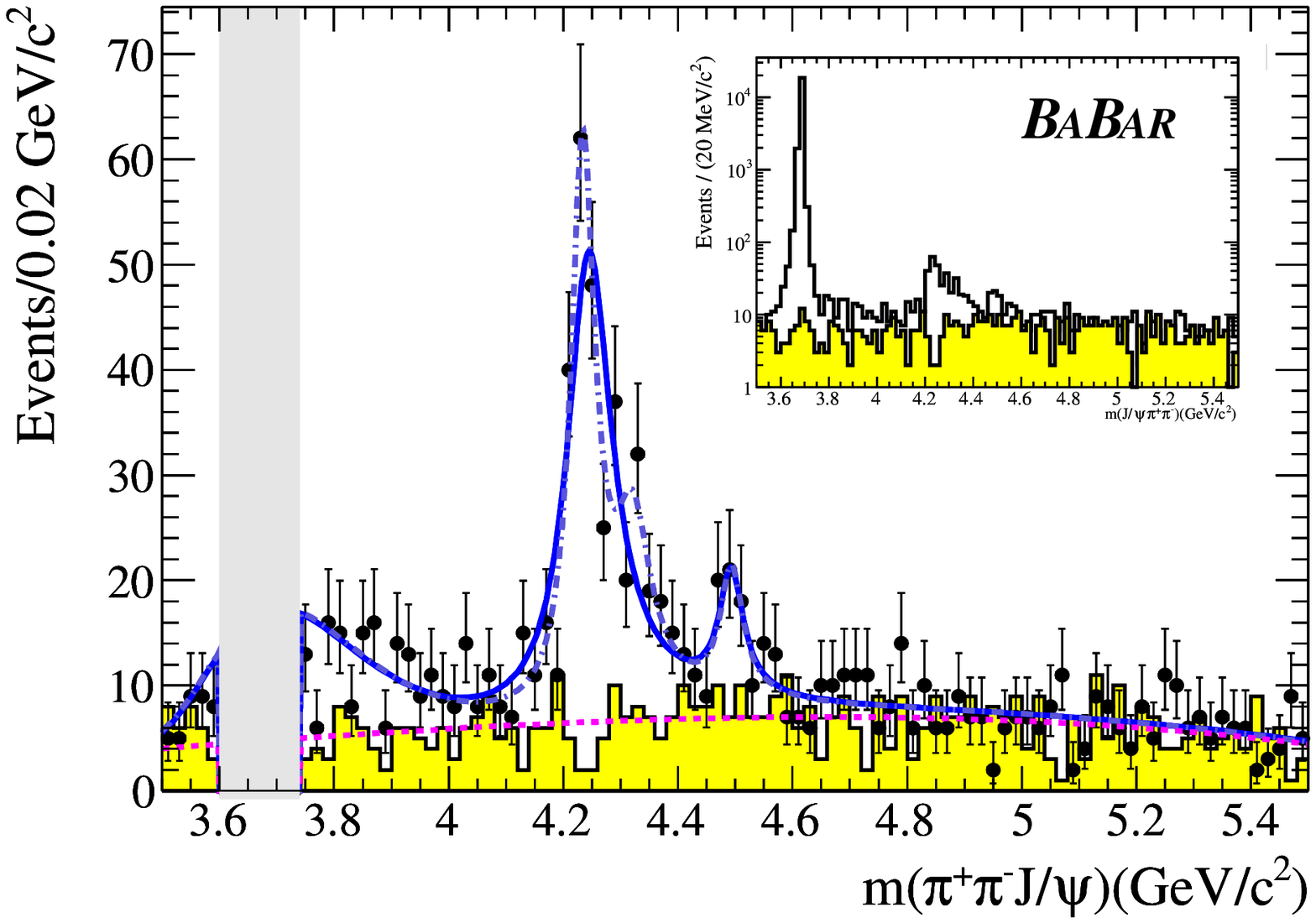}
\caption{The $J/\psi \pi^+\pi^-$  mass spectrum in the  
$e^+e^-\to J/\psi \pi^+\pi^-\gamma$ reaction.}
\label{4260}
\end{minipage}
\hfill
\begin{minipage}[t]{0.30\textwidth}
\includegraphics[width=.98\textwidth]{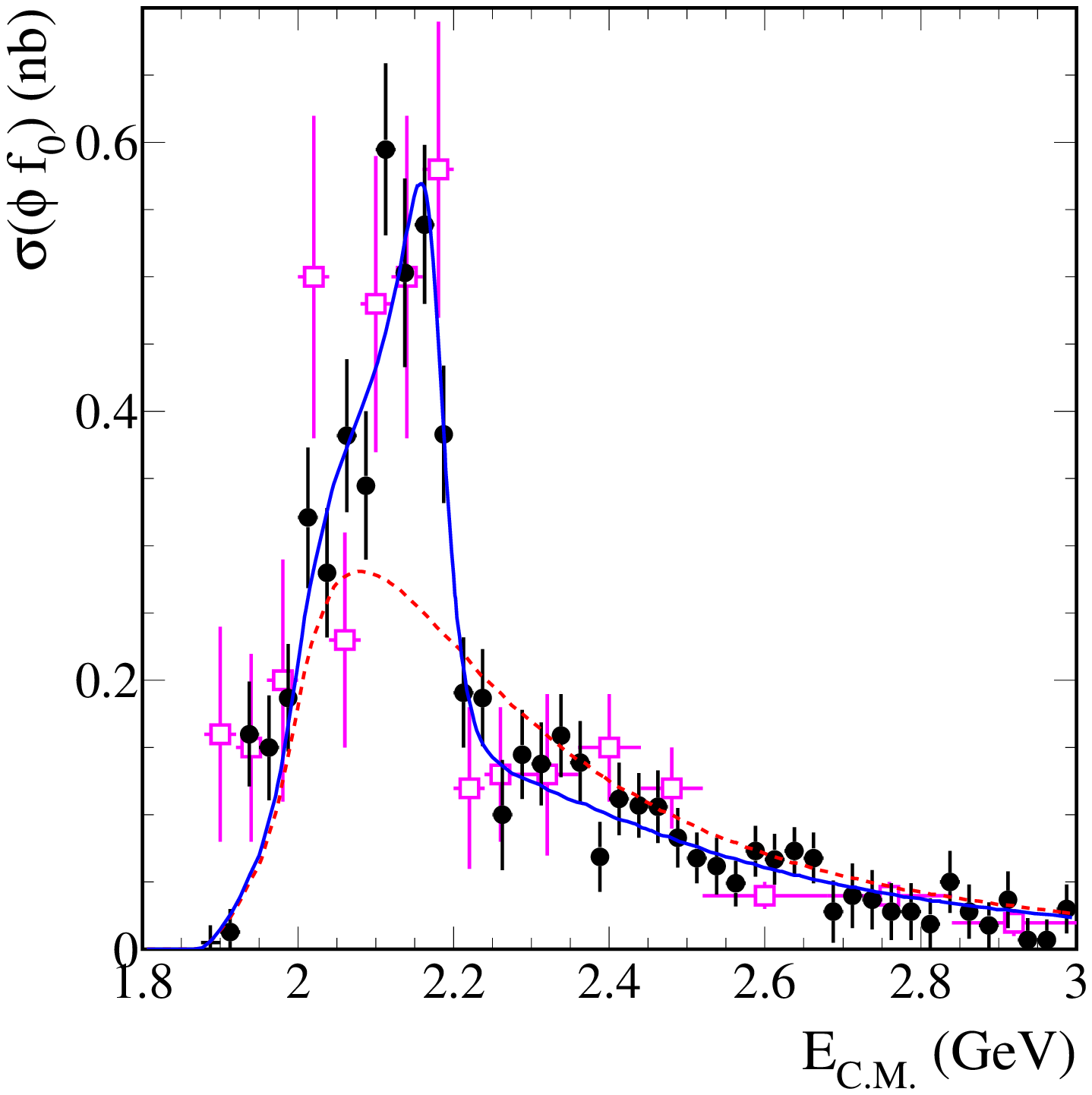}
\caption{ The  \babar\ $e^+e^-\to \phi f_0$  cross section,
measured in $K^+K^-\pi^+\pi^-$ (filled circles) and
$K^+K^- \pi^0\pi^0$ (squares) final states. }
\label{2175}
\end{minipage}
\end{figure}

{\underline{\bf Production of baryons }}
   The first  \babar\ ISR baryon experiment was the  measurement
   of the proton electromagnetic form
factor (FF) \cite{BBprot}  
in the reaction $e^+e^-\to p\bar{p}\gamma$.
The  \babar\ proton FF data, presented in  Fig.\ref{prot1}, in general agree
with the previous measurements,  confirm the rise
of the proton FF at the threshold and  modestly satisfy the QCD fit formula  
\cite{Chern2}  above $~$ 2.5 GeV.
Structures  at 2.15 and 2.9 GeV are  possibly seen. Study of the angular
distribution allows to extract the value of $|G_E/G_M|$ (the ratio of the 
electric
and magnetic FFs) (Fig.\ref{prot2}),  which is found to be contradicting
the only previous measurement at LEAR \cite{LEAR}.

    Later on the reactions $e^+e^-\to \Lambda\bar{\Lambda},~
\Sigma\bar{\Sigma},~ \Lambda\bar{\Sigma}(\Sigma\bar{\Lambda})$ 
have been studied  \cite{BBLS}
and corresponding FFs are measured as well. The most accurate FF measurement is
done for $\Lambda$. Previously the  $\Lambda$ FF was measured by   DM2
only \cite{DM2LL} in a single  energy point. 
The  \babar\ data on the $\Lambda$ and proton FF are
shown in Fig.\ref{prot3}. The asymptotic QCD prediction \cite{Chern}
for the ratio
of the $\Lambda$ and proton FF is 0.24. Figure \ref{prot3} shows the strong
disagreement between the  data and prediction below 2.5 GeV. The disagreement at 
higher masses is less, but the measurements accuracy is not sufficient for
the test of the prediction.

{\underline{\bf Resonance physics}}
  The number of produced events   
  of narrow resonance with  ISR is described by the
expression $N_{res}=\sigma_{res}\cdot L$, where 
$\sigma_{res}=\frac{12\pi^2\Gamma_{ee}B_f}{s~m_{res}}\cdot
W(s,x_0,\theta^*_0)$ is the ISR cross section
and $L$ is the total 
integrated luminosity. The $x_0$ is related to the mass $m_0$  
of the resonance:  $x_0=1 -m_0^2/s $.
In the expression for the ISR cross section
the $\Gamma_{ee},~B_f$ are the electron width and decay branching
fraction, respectively. 
One can conclude from the above expression, that the number of events  
is determined by  the electron width of the resonance. 

     Using the ISR technique, parameters of known vector states $\rho (1450),
~\rho (1700)$, $\omega (1420)$, $\omega (1650)$ and 
 $\phi (1650)$ were improved (see above references on the 
 cross section measurements).
 Concerning the $J/\psi$ and $\psi (2S)$ resonances,
many  decays 
were observed for the first time ($J/\psi\to K^+K^-\pi^0\pi^0,$
 $K^+K^-\pi^+\pi^-\eta$, $ K^{*0}\bar{K^{*0}}$, $\phi\pi^0~ {\ldots} $, 
$\psi (2S)\to \pi^+\pi^-\pi^+\pi^-\eta$,  $K^+K^-\pi^+\pi^-\pi^+\pi^-~ {\ldots}
$),   and the accuracies  of  many known  decays were improved.

     New vector states, found in ISR, deserve special attention. The most known
is the charmonium-like Y(4260) resonance \cite{Y4260}, 
decaying into $J/\psi \pi^+\pi^-$.  It is relatively narrow (Fig.\ref{4260}), 
$\Gamma = 92\pm 15~MeV/c^2$,  and has very low electron width 
$\Gamma_{ee}=7.5\pm 1.1~eV$, which is much less than the   value
$\sim500-800~eV$,  typical for  $\psi (4040),\psi (4160),\psi (4400)$ states. 
No decays of Y(4260) into $D\bar{D}$ is seen. 
For these reasons the Y(4260) is considered as      
a candidate for the exotic 4-quark $c\bar{c}n\bar{n}$ ($n=u,d$) or molecular
 state. One could mention, that the $\pi^+\pi^-$ mass spectrum 
in Y(4260) decay 
has a peak close to 1 GeV.  This  might be the indication of the 
contribution of $J/\psi f_0(980)$ intermediate state.
   
  Another new state is  Y(4320) \cite{Y4320},
decaying into $\psi (2S)\pi^+\pi^-$.  Judging by presence
of charmonium in the final state, narrow width and small production cross
section this state might be of a similar nature as Y(4260).
   
   One more  exotic candidate is the Y(2175) \cite{Y2175},   
   first seen by  \babar\ in
$K^+K^-f_0(980)$, $f_0\to \pi^+\pi^-,\pi^0\pi^0$ final states
(Fig.\ref{2175})
and then in the $\phi\eta$ (Fig.\ref{fieta}) state. Similar to Y(4260) it
is relatively narrow, has very low electron width and is considered
as a possible $s\bar{s}s\bar{s}$ exotic candidate or
$\phi^{\prime\prime}$    state.
      
      Summarising the data on isovector channels one can suggest the
existence of the new isovector state $\rho (2150)$ or
$\rho^{\prime\prime\prime}$.
This resonance is seen in $\pi^+\pi^-\pi^0\pi^0$  (Fig.\ref{4pi2}), 
$\eta^{\prime}(958)\rho^0$  \cite{BB5pi}  and
$p\bar{p}$ final states. 
In the latter case, we mean the step in the proton form factor near
2.15 GeV (Fig.\ref{prot1}). 

     One should mention also about one more isovector candidate X(1880),  
seen by  \babar\ as interference pattern (Fig.\ref{6pi}) in the $e^+e^-\to 6\pi$ 
 cross section \cite{BB6pi}. This state was seen earlier
in DM2 \cite{DM26pi}   and FOCUS \cite{FOCUS6pi}  experiments. 
The enhancement in the proton form factor
at the threshold  (Fig.\ref{prot1})  can be due to  this X(1880) state too.

{\underline{\bf Conclusions}}
     The $e^+e^-\to hadrons$ cross sections 
are measured at  \babar\   in the energy range $1\div 5$ GeV
using the ISR method.
Parameters of many vector mesons
($\rho, \omega, \phi$ excitations and the $J/\psi$ and $\psi (2S)$ states)
are improved.
New resonances are observed via ISR: Y(4260), Y(4320), and Y(2175),
which are candidates for exotic states.  Also seen are two new 
isovector candidates  $\rho (2150)$ and X(1880).



\end{document}